# Polarization-resolved terahertz third-harmonic generation in a superconductor NbN: dominance of Higgs mode beyond the BCS approximation

Ryusuke Matsunaga[1,2], Naoto Tsuji[3], Kazumasa Makise[4], Hirotaka Terai[4], Hideo Aoki[1,5], and Ryo Shimano[1,6]

[1]*Department of Physics, The University of Tokyo, Hongo, Tokyo, 113-0033, Japan*
[2]*PRESTO, Japan Science and Technology Agency, 4-1-8 Honcho Kawaguchi, Saitama 332-0012, Japan*
[3]*RIKEN Center for Emergent Matter Science, Wako, Saitama 351-0198, Japan*
[4]*National Institute of Information and Communications Technology, 588-2 Iwaoka, Kobe 651-2492, Japan*
[5]*Electronics and Photonics Research Institute, Advanced Industrial Science and Technology, Umezono, Tsukuba, Ibaraki 305-8568, Japan*
[6]*Cryogenic Research Center, The University of Tokyo, Yayoi, Tokyo, 113-0032, Japan*

**Abstract**

**Recent advances in time-domain terahertz (THz) spectroscopy have unveiled that resonantly-enhanced strong THz third-harmonic generation (THG) mediated by the collective Higgs amplitude mode occurs in *s*-wave superconductors, where charge-density fluctuations (CDF) have also been shown to contribute to the nonlinear third-order susceptibility. It has been theoretically proposed that the nonlinear responses of Higgs and CDF exhibit essentially different polarization dependences. Here we experimentally discriminate the two contributions by polarization-resolved intense THz transmission spectroscopy for a single-crystal NbN film. The result shows that the resonant THG in the transmitted light always appears in the polarization parallel to that of the incident light with no appreciable crystal axis dependence. When we compare this with the theoretical calculation here with the BCS approximation and the dynamical mean-field theory for a model of NbN constructed from first principles, the experimental result strongly indicates that the Higgs mode rather than the CDF dominates the THG resonance in NbN. A possible mechanism for this is discussed such as the retardation effect in the phonon-mediated pairing interaction beyond BCS.**

*Introduction* --- Spontaneous symmetry breakdown in many-body systems has been one of the central interests in condensed matter physics. Collective excitations arising from fluctuations of amplitude and phase of an order parameter are particularly intriguing as an inherent manifestation of a symmetry breaking, which have recently attracted renewed interests [1, 2] since experimental techniques for accessing these modes were developed with ultrafast spectroscopy [3-6] or artificial control of physical parameters in the vicinity of quantum critical points [7-10]. Especially, the amplitude mode of the order parameter in superconductors has a close analogy with the Higgs boson in particle physics [11, 12], hence called the Higgs amplitude mode [1].

The Higgs mode in superconductors has been identified with a Raman spectroscopy in $NbSe_2$, where a coexisting charge-density wave makes the mode Raman-active [13-17]. For ordinary superconductors, however, the Higgs mode has eluded experimental detection until recently [18-20] primarily because the Higgs mode does not couple to electromagnetic fields in the linear-response regime [21]. Recently, a Higgs mode oscillation with the superconducting gap frequency $2\Delta$ was directly observed in a conventional *s*-wave superconductor $Nb_{1-x}Ti_xN$ by a terahertz (THz) pump-THz probe experiment [18] as the oscillation of order parameter in time domain after a nonadiabatic excitation [22-28]. Subsequently, it was revealed that irradiation of an intense narrow-band THz wave onto NbN with the photon energy $\omega$ tuned below $2\Delta$ induces a third-harmonic generation (THG) [19]. A salient feature is that the THG intensity is strongly enhanced when the incident frequency doubled, $2\omega$, coincides with the gap, $2\Delta$. While THG from superconductors has been discussed phenomenologically in terms of a nonlinear supercurrent model [29], the resonant enhancement of the THG at $2\omega=2\Delta$ has revealed the existence of a *nonlinear coupling* between the Higgs mode and electromagnetic wave [24, 30]. Such a nonlinear THz spectroscopy provides a new tool for studying the collective modes which are now being theoretically studied for other types of exotic superconductors, *e.g.*, multi-band [31-34] or *d*-wave superconductors [35].

Importantly, it has been pointed out that in addition to the Higgs mode, the charge-density fluctuation (CDF) or pair breaking, which has conventionally been identified as the origin of the peak at $2\Delta$ in Raman spectroscopy [36], also induces the THG with a similar resonant character at $2\omega=2\Delta$ [37]. Within the BCS mean-field approximation the CDF contribution is shown to be typically much larger than the Higgs mode contribution [37]. However, their relative magnitude should depend sensitively on how we take account of the many-body interactions. Indeed, a recent calculation with the dynamical mean-field theory (DMFT) has revealed that the BCS approximation significantly underestimates the Higgs-mode contribution because some of the important diagrams for the nonlinear optical susceptibility accidentally vanish in the BCS framework [38]. If we take account of dynamical correlations such as the retarded electron-phonon coupling or impurity scattering, the contribution of the Higgs mode to the THG is shown to be significantly enhanced and can even exceed the CDF [38]. NbN is in fact a strongly electron-phonon-coupled system with the dimensionless coupling constant $\lambda \sim 1$ [39-41], for which the retardation effect can invalidate the weak-coupling BCS treatment. It is thus imperative to decompose the Higgs and CDF contributions by experiments. One promising key is the dependence of the nonlinear susceptibility on the direction of the electric field polarization of the laser with respect to crystal axes of a superconductor, as discussed for a square lattice in Ref. [37].

The purpose of this paper is to discriminate the Higgs mode and the CDF contributions by the polarization dependence of the nonlinear THG response for a conventional superconductor NbN. We theoretically evaluate the CDF contribution to the THG for the three-orbital model of NbN constructed from first principles using the BCS approximation and DMFT. We shall show that the CDF contribution to THG increases by a factor of 2.3-3.1 when the polarization angle changes from [100] to [110]. We also find that the CDF contains a polarization component perpendicular to the incident field polarization. The Higgs contribution, on the other hand, always appears parallel to the incident one with no angle dependence, as generally proved by the symmetry argument. Experimentally, we shall show in a polarization-resolved intense THz transmission spectroscopy for a single-crystal NbN film that the THG polarization is indeed parallel to the incident field and that the THG intensity hardly changes against the crystal axis orientation. From these theoretical and experimental results, we shall conclude that the Higgs mode plays a dominant role in the resonant enhancement of THG around $2\omega=2\Delta$ in NbN.

*Theoretical analysis* --- We construct an effective low-energy model of NbN based on a first-principles density-functional calculation using the WIEN2k package [42]. In Fig. 1(a) we display the band structure of NbN, which agrees with the previous results [43-45]. There are three bands around the Fermi energy ($E=0$), which are mainly composed of $4d$ $t_{2g}$ orbitals ($d_{xy}$, $d_{yz}$ and $d_{zx}$) of Nb. We illustrate $4d_{xy}$ orbitals on the fcc lattice in Fig. 1(b). The neighboring bands coming from Nb $4d$ $e_g$ and N $2p$ orbitals are well separated from the $t_{2g}$ bands in energy by a few eV [Fig. 1(a)], which motivates us to build an effective three-band tight-binding model in terms of the $t_{2g}$ orbitals.

For each orbital we take three hopping processes with amplitudes $t$, $t'$, and $t''$ as shown in Fig. 1(b). Since different $t_{2g}$ orbitals at neighboring sites are orthogonal to each other, inter-orbital hoppings are suppressed. The resulting Hamiltonian reads

$$H_0 = \sum_{k,a,\sigma} \varepsilon_a(k) d_{a,\sigma}^\dagger(k) d_{a,\sigma}(k),$$

where $a=xy$, $yz$, $zx$ labels the orbitals, $d_{a,\sigma}^\dagger(k)$ creates a $d$-electron with orbital $a$, spin $\sigma$, and momentum $k$, and

$$\varepsilon_{xy}(k) = 4t \cos\frac{k_x}{2} \cos\frac{k_y}{2} + 2t'(\cos k_x + \cos k_y) + 4t''\left(\cos\frac{k_y}{2}\cos\frac{k_z}{2} + \cos\frac{k_z}{2}\cos\frac{k_x}{2}\right),$$

with $\varepsilon_{yz}(k)$ and $\varepsilon_{zx}(k)$ given by permuting $x$, $y$, $z$. We fit the band dispersion with the result of the first-principles calculation to obtain the hopping parameters as $t=-0.72$ eV, $t'=-0.15$ eV, $t''=0.12$ eV, and the chemical potential $\mu=-0.6$ eV. The band dispersion of the effective model, plotted in red in Fig. 1(a), shows that the $t_{2g}$ bands are well reproduced by the effective model around the Fermi energy.

The polarization dependence of the THG is evaluated in the BCS approximation and in the DMFT. For the BCS, we take the pairing interaction,

$$H_{\text{int}} = -\frac{1}{N} \sum_{k,k',abcd} v_{ad,cb}(k,k') d_{a\uparrow}^\dagger(k) d_{b\downarrow}^\dagger(-k) d_{c\downarrow}(-k') d_{d\uparrow}(k'),$$

where $N$ is the number of $k$ points and $v_{ad,cb}(k,k')$ is the scattering matrix element, which can be expanded in each sector of irreducible representations $\Gamma$ of the point group ($O_h$) for NbN as $v_{ad,cb}(k,k') = \sum_\Gamma \sum_i v^\Gamma [\hat{\varphi}_i^\Gamma(k)]_{ab} [\hat{\varphi}_i^\Gamma(k')]_{dc}^*$. Here $v^\Gamma$ is the interaction parameter for sector $\Gamma$ with $\hat{\varphi}_i^\Gamma(k)$ being its $i$th basis function. We assume that the superconducting pairing realized in NbN belongs to the spin-singlet and orbital $A_{1g}$ representation (with $[\hat{\varphi}_i^\Gamma(k)]_{ab} \propto \delta_{ab}$), and neglect the effect of the pairing interactions other than the $A_{1g}$ sector. To reproduce the experimental condition for NbN, we take the model parameters for the superconducting gap $\Delta=2.7$ meV$=0.65$ THz and the temperature $T=4$ K$=0.34$ meV. In DMFT, we consider a three-orbital Holstein model,

$$H = \sum_{k,a,\sigma} \varepsilon_a(k) d_{a,\sigma}^\dagger(k) d_{a,\sigma}(k) + \omega_0 \sum_i a_i^\dagger a_i + g \sum_{i,a} \left(a_i^\dagger + a_i\right)\left(n_{ia} - \langle n_{ia}\rangle\right),$$

where $a_i^\dagger$ creates a phonon at $i$ with frequency $\omega_0$, $g$ is the electron-phonon coupling, and $n_{ia} = \sum_\sigma d_{ia\sigma}^\dagger d_{ia\sigma}$ is the electron density. The impurity problem for DMFT is solved by the (unrenormalized) Migdal approximation. The parameters are taken to be $\omega_0$=2.0 eV, $g$=2.0 eV, and $T$=0.05 eV as an example. We have confirmed that the results do not qualitatively change as the parameters are varied. The method for the calculation of the THG susceptibility is summarized in Supplemental Material [46]. We set the polarization $e^I$ of the incident light and the polarization $e^O$ along which the transmitted light is probed to be $e^I = e^O = e_\theta = (\cos\theta, \sin\theta, 0)$.

We plot the BCS and DMFT results for the THG intensity $|\chi(\omega)|^2$ for $\theta$=0° and 45° in Fig. 2(a)(b). One can see that the CDF contribution has a resonance at $2\omega=2\Delta$ for each $\theta$. The intensities of the CDF at the resonance $|\chi(2\omega=2\Delta)|^2$, normalized by the value at $\theta$=0°, are plotted against $\theta$ in Fig. 2(c)(d). The CDF contribution has a characteristic polarization dependence with its intensity increasing by a factor of 2.3 (3.1) as $\theta$ is varied from 0° to 45° in the BCS (DMFT) result. On the other hand, the intensity of the Higgs-mode contribution does not depend on $\theta$ [Fig. 2(c)(d)]. Although the relative magnitude between the CDF and Higgs is very different between the BCS and DMFT results [38], the polarization dependence of the THG is qualitatively similar between the BCS and DMFT. The results for the polarization dependence of the CDF can be qualitatively understood as follows: If we neglect the subleading $t'$ and $t''$ hoppings for simplicity, the tight-binding model on the fcc lattice consists of a set of two-dimensional square lattices on $xy$, $yz$, and $zx$ planes respectively but rotated by 45° on each plane [Fig. 1(c)]. It has been shown [37] that for a square lattice the CDF is maximally enhanced (suppressed) for $\theta$=0° ($\theta$=45°), which implies for the present case that the CDF is enhanced (suppressed) for $\theta$=45° ($\theta$=0°). The result in Fig. 2(c)(d) indicates that the corrections due to $t'$ and $t''$ do not significantly change the polarization dependence.

A general form for the polarization dependence of the CDF and Higgs contributions to the THG susceptibility $\chi(\omega)$ is given in Table 1. One can see that the CDF contribution is non-vanishing when $e^O$ is perpendicular to $e^I$. For a general band dispersion, $A(\omega)$ and $B(\omega)$ in Table 1 have similar orders of magnitude. This means that, if the CDF contribution is dominant, the THG should also be observed for the direction perpendicular to the polarization of the incident light. This sharply contrasts with the Higgs-mode contribution, which appears only in the direction parallel to the polarization of the incident light, and does not depend on $\theta$. This can be generally understood by the symmetry argument [46] based on the fact that the $s$-wave pairing is isotropic in the momentum and orbital spaces.

*Experimental analysis* --- We measured the dependence of THG on the electric field polarization with respect to lattice axes. The sample is a NbN thin film on a MgO substrate [47] with $T_c$ = 15 K. From X-ray scattering we confirmed that the single-crystal (100) NbN was epitaxially grown on (100) MgO with cube-on-cube in-plane alignment [46]. Figure 3(a) shows a schematic experimental setup for polarization-resolved THz transmission spectroscopy. Strong monocycle THz pulses with vertical polarization (//$x$ or 0º) were generated by optical rectification in a LiNbO$_3$ crystal with the tilted-pulse-front scheme [48-50]. Bandpass filters were placed to make narrow-band THz pulses with the center frequency of $\omega$=0.5 THz. In front of the sample we set two wire-grid polarizers WGP1 and WGP2, whose angles $\theta_1$ and $\theta_2$, respectively, determine the field strength as factored by $\cos\theta_1\cos(\theta_2-\theta_1)$. The electric field polarization on the sample is determined solely by $\theta_2$. Angles $\theta_2$=0º and 45º correspond to [100] and [110] directions, respectively, as indicated in the inset of Fig. 3(a). Additional two polarizers WGP3 and WGP4 are placed behind the sample with angles $\theta_3$ and $\theta_4$, respectively. The WGP3 is set to $\theta_3=\theta_2$ or $\theta_2$+90º for detection of the THG polarized parallel or perpendicular to the incident field, respectively. Extinction ratio for this set up was evaluated as ~10$^{-4}$ in the frequency range below 3$\omega$=1.5 THz [46], which is good enough for resolving the polarization state of THG. Transmitted THz pulses were detected by the electro-optic (EO) sampling with a (100) ZnTe crystal.

We first examine the nonlinear transmission spectra in the case of $\theta_2$=22.5º, where the CDF should give rise to the THG polarized perpendicular to the incident field according to Table 1. Here we set $\theta_4=\theta_2$+45º so that the transmitted electric field parallel ($\theta_3=\theta_2$) or perpendicular ($\theta_3=\theta_2$+90º) to the incident field can be directly compared because both are detected with the same projection of 45º on the WGP4. Figure 3(b) shows the power spectra of the transmitted pulse with the peak electric field of $E_{THz}$~5 kV/cm. The black curve shows the data above $T_c$. The red and blue curves correspond to the parallel and perpendicular configurations, respectively, at $T$=11.5 K < $T_c$ at which 2$\omega$=2$\Delta(T)$ is satisfied. For the parallel configuration THG is clearly observed at 3$\omega$=1.5 THz, in a stark contrast to the perpendicular configuration where no THG signal was identified. For other incident polarization angles we observe no THG signals for the perpendicular configuration, either. The THG component parallel to the incident polarization is at least 10$^3$ times larger than the perpendicular one, which means that |$B(\omega)$| in Table 1 is much smaller than max{|$A(\omega)$|, |$C(\omega)$|}.

We also investigate the dependence of the THG intensity on the incident field polarization direction by rotating $\theta_2$ from 0º to 45º. For each $\theta_2$, WGP1 is tuned so as to fix $\cos\theta_1\cos(\theta_2-\theta_1)$=0.85, hence the field strength constant. WGP3 is also rotated as $\theta_3=\theta_2$ to detect the THG polarized parallel to the incident field. WGP4 is fixed at $\theta_4=45°$ to maintain the field polarization detected by the EO sampling. From the power spectra we obtained the observed THG intensity $I_{3\omega}^{\text{obs}}$ and the observed fundamental intensity $I_{\omega}^{\text{obs}}$. Note that the observed values of $I_{3\omega}^{\text{obs}}$ and $I_{\omega}^{\text{obs}}$ are related with the generated THG intensity $I_{3\omega}(\theta_2, E_{\text{THz}})$ and the transmitted fundamental intensity $I_{\omega}(\theta_2, E_{\text{THz}})$, respectively, by a factor of $\cos^2(\theta_2\text{-}45°)$ because of the projection on the WGP4. Thus we focused on the ratio $R(\theta_2) = I_{3\omega}^{\text{obs}} / I_{\omega}^{\text{obs}} = I_{3\omega}(\theta_2, E_{\text{THz}}) / I_{\omega}(\theta_2, E_{\text{THz}})$ to cancel out the projection factor. $R(\theta_2)$ is then proportional to $|\chi(\theta_2)|^2 |E_{\text{THz}}|^4$, where $\chi(\theta_2)$ is the third-order nonlinear susceptibility. We also checked the field strength $E_{\text{THz}}$ during the rotation of the WGPs and confirmed that fluctuation of $E_{\text{THz}}$ is negligibly small [46]. Then we obtained the squared nonlinear susceptibility $|\chi(\theta_2)|^2 \propto R(\theta_2)/|E_{\text{THz}}(\theta_2)|^4$, as displayed in Fig. 3(c) where the data is normalized at $\theta_2=0°$. The THG intensity is seen to be basically constant, changing only within 5±6% from [100] to [110] directions, namely, the THG intensity hardly depends on the incident field polarization with respect to the crystal axis. Since the polarization angle dependence of THG arises only from $B(\omega)$ in Table 1, the experimental result in Fig. 3(c) elucidates that $|B(\omega)|$ is much smaller than $\max\{|A(\omega)|, |C(\omega)|\}$, which is consistent with the experimental result in Fig. 3(b).

Because the calculations in Figs. 2(c) and 2(d) indicate that $|A(\omega)|$ and $|B(\omega)|$ are of the same order of magnitude, we can conclude that $|C(\omega)| \gg |A(\omega)|, |B(\omega)|$, which means that the contribution of Higgs mode to the THG is much larger than the CDF. This is intriguing, since the result for the relative magnitudes of the two contributions is opposite to the BCS prediction [37]. A possible mechanism for the dominance of the Higgs-mode contribution is that the THG process beyond the BCS approximation contains the retardation effect that significantly enhances the contribution of the resonant THG diagram in strongly electron-phonon-coupled superconductors [38]. By taking account of a relationship between the THG and Raman process, where the latter probes the imaginary part of the third-order susceptibility [51], it is worth noting that our results imply that the resonant diagrams may play non-negligible role also in the Raman process in systems with strong retardation effects [52].

*Summary* --- We have studied the polarization dependence of the THG in a superconductor NbN theoretically and experimentally, and revealed that the Higgs mode gives a dominant contribution to the THG far exceeding the CDF contribution. The results also demonstrate that the polarization-resolved nonlinear THz spectroscopy provides a new pathway for investigating collective modes in superconductors. An important future problem is to extend the present scheme to unconventional superconductors such as the high-$T_\text{c}$ cuprates.

We wish to thank Y. Gallais for illuminating discussions. This work was supported in part by JSPS KAKENHI (Grants Nos. JP15H05452, JP15H02102, JP26247057, and JP16K17729), by the Photon Frontier Network Program from MEXT, Japan, by PRESTO, JST, and by ImPACT project (Grant No. 2015-PM12-05-01).

**Figure captions**

**Fig. 1** (Color online) (a) Band structure of NbN obtained from a first-principles calculation with the weights of the band character of Nb 4d $e_g$, 4d $t_{2g}$, and N 2p displayed in green, orange, and blue, respectively. Red curves represent the band in the effective three-orbital model. (b) Nb $4d_{xy}$ orbitals on the fcc lattice. The arrows represent the hoppings with amplitudes *t*, *t'*, and *t"*. (c) Sets of two-dimensional square lattices made of $d_{xy}$, $d_{yz}$, and $d_{zx}$ orbitals respectively on *xy*, *yz*, and *zx* planes with a rotation of 45° on each plane.

**Fig. 2** (Color online) (a), (b) The CDF contribution to the THG intensity spectra $|\chi(\omega)|^2$ with $\boldsymbol{e}^I = \boldsymbol{e}^O = \boldsymbol{e}_\theta$ for *θ*=0° (lower curve) and 45° (upper) calculated by the BCS approximation (a) and DMFT (b). (c), (d) The polarization dependence of the CDF and Higgs-mode contributions to the THG intensity $|\chi(\omega)|^2$ at the resonance (2*ω*=2Δ) with $\boldsymbol{e}^I = \boldsymbol{e}^O = \boldsymbol{e}_\theta$ calculated by the BCS approximation (c) and DMFT (d). The intensity in each panel is normalized by the value at *θ*=0°. For the BCS approximation we take Δ=2.7 meV=0.65 THz and *T*=4 K, and for DMFT we take the Holstein model with $\omega_0$=2.0 eV, *g*=2.0 eV, and *T*=0.05 eV.

**Fig. 3** (Color online) (a) A schematic experimental setup for THz transmission spectroscopy. WGP: wire-grid polarizer, BPF: band-pass filter. Inset shows the electric field polarization along the crystal axis on the sample surface. (b) Experimental result for the power spectra of the transmitted THz pulse with $\theta_2$=22.5°. The black curve is obtained above $T_c$. Red and blue curves show the data at 2*ω*=2Δ(*T*) for polarizations parallel and perpendicular to the incident field, respectively. (c) Squared nonlinear susceptibility $|\chi|^2$, normalized at $\theta_2$=0°, as a function of the incident polarization angle $\theta_2$.

**Table 1** The polarization dependence of the THG susceptibility $\chi(\omega)$ relevant to the resonance at 2*ω*=2Δ for the CDF and Higgs-mode contributions. The polarization-independent functions *A*(*ω*), *B*(*ω*), and *C*(*ω*) are defined in [46].

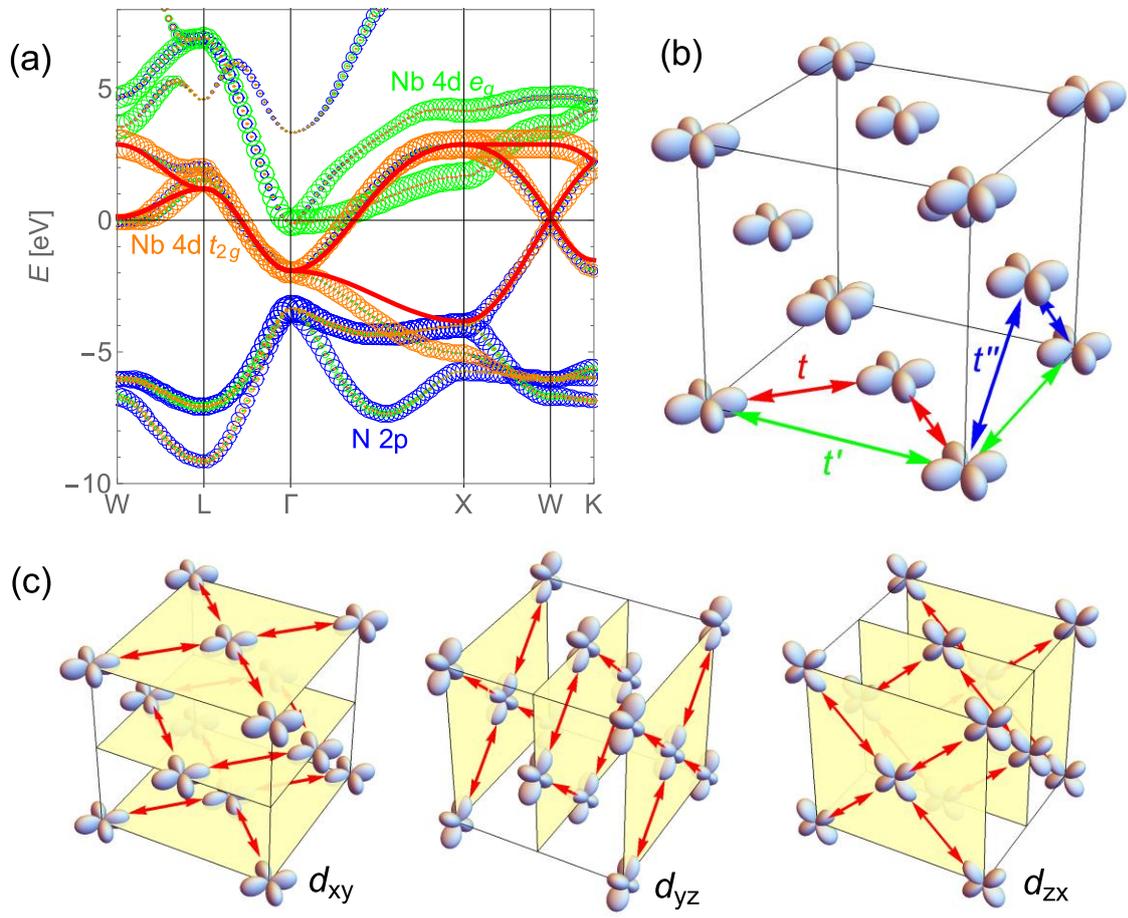

Figure 1

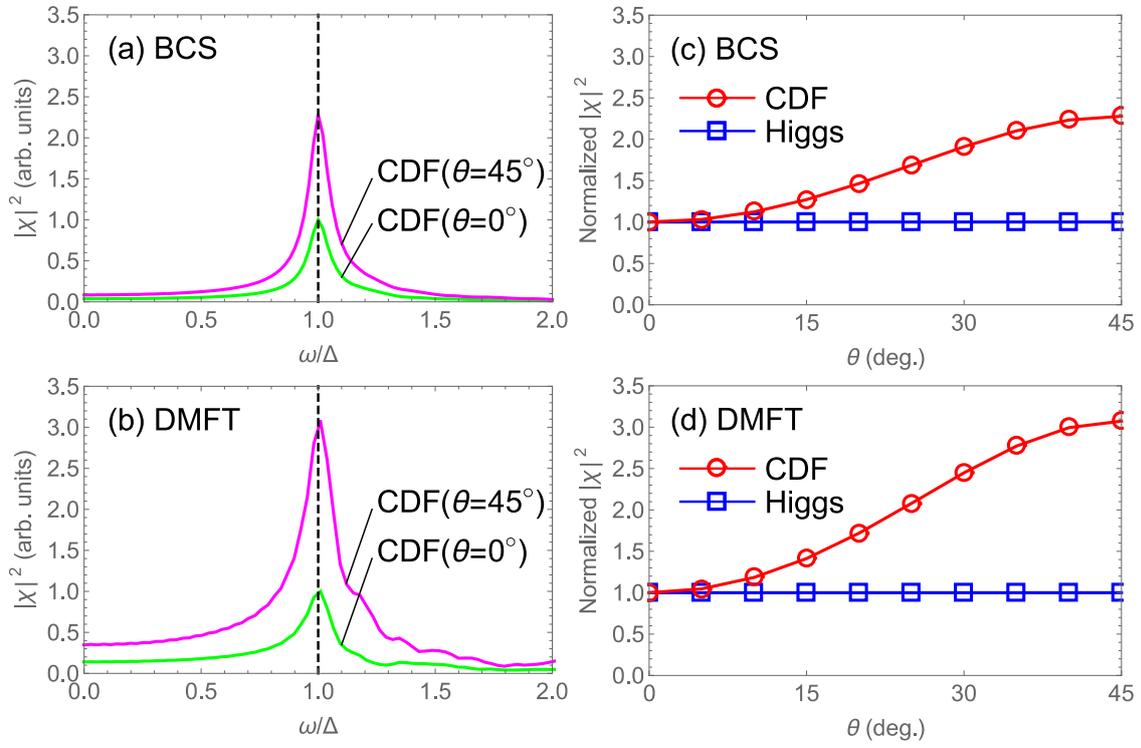

Figure 2

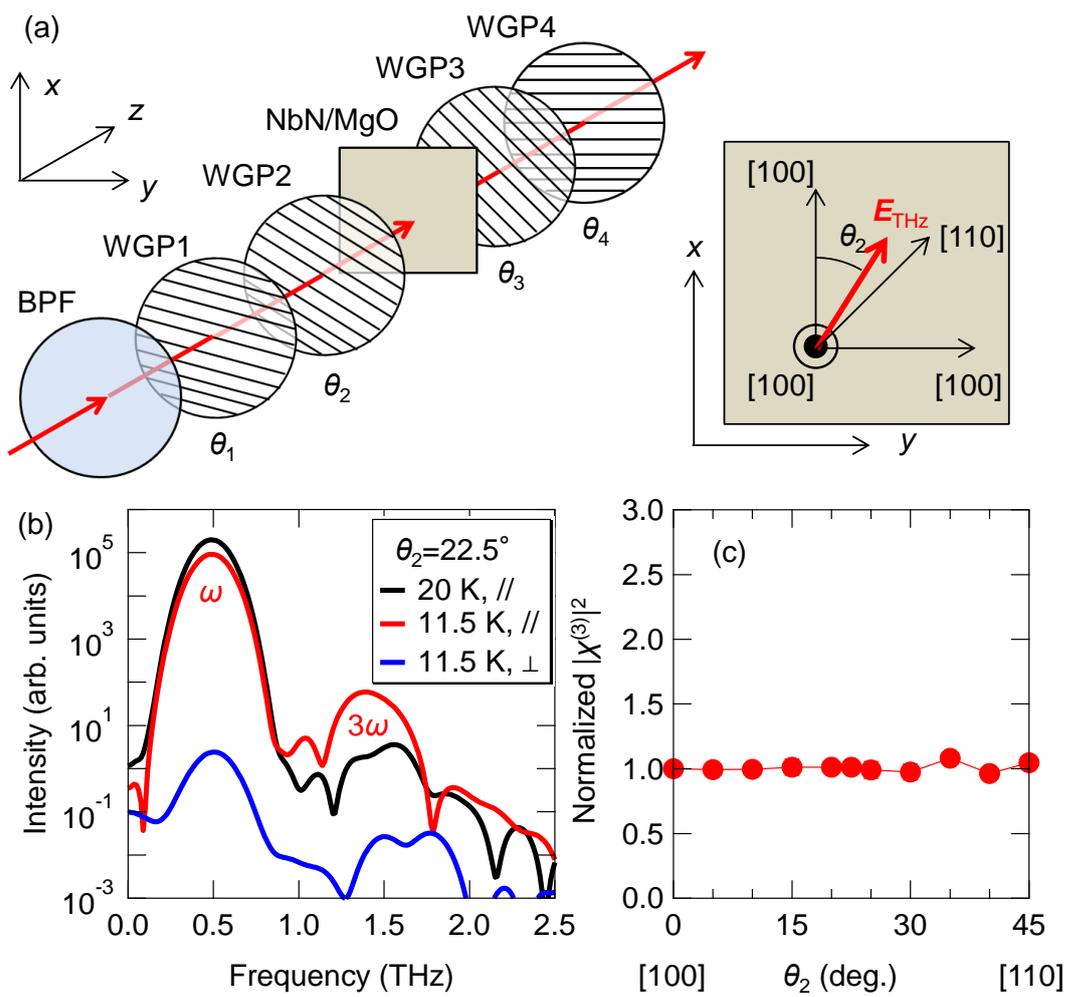

Figure 3

|  | $e^I = e^O = e_\theta$ | $e^I = e_\theta,\ e^O = e_{\theta+90°}$ |
|---|---|---|
| CDF | $A(\omega)+2B(\omega)\sin^2 2\theta$ | $B(\omega)\sin 4\theta$ |
| Higgs | $C(\omega)$ | 0 |

Table 1

**Supplemental Material for**

**Polarization-resolved terahertz third-harmonic generation in a superconductor NbN: dominance of Higgs mode beyond the BCS approximation**

Ryusuke Matsunaga, Naoto Tsuji, Kazumasa Makise, Hirotaka Terai, Hideo Aoki, and Ryo Shimano

**Sample Characterization**

The sample is a 24-nm NbN thin film fabricated on a (100) single-crystal MgO substrate with 500-µm thickness with the dc reactive sputtering method [S1]. Figure S1(a) shows X-ray diffraction traces of the sample with the $\theta$-$2\theta$ scan method. Strong (200) diffraction peaks from both the NbN thin film and the MgO substrate were observed. Figures S1(b) and S1(c) show the X-ray diffraction patterns with the $\Phi$-scan method. Under 360º rotation the [220] peaks of the NbN film appears at the same angles as the [220] peaks of the MgO substrate. These confirm that single-crystal [100] NbN was epitaxially grown on the [100] MgO substrate with excellent cube-on-cube in-plane alignment.

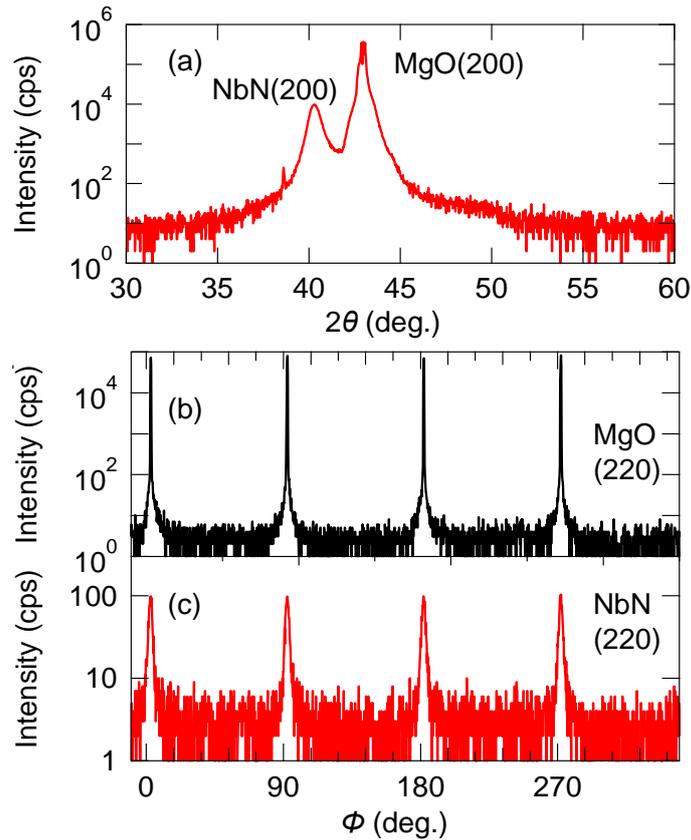

Fig. S1. X-ray diffraction patterns of the NbN/MgO sample with (a) the $\theta$-$2\theta$ method and (b) the $\Phi$-scan method, respectively.

**Experimental Setup and Polarization Resolution**

Figure S2(a) shows the experimental setup for polarization-resolved intense terahertz (THz) transmission spectroscopy. Output from a Ti:Sapphire-based regenerative amplified laser, with the pulse energy of 1.2 mJ, the center wavelength of 800 nm, the pulse duration of 100 fs, and the repetition rate of 1 kHz, was divided for the THz pulse generation and detection. The main pulse of the laser output was irradiated on a $LiNbO_3$ crystal after tilting the pulse front by grating [S2-S4] to generate the strong monocycle-like THz pulse. A black polypropylene film was inserted in the THz beam path to block the remnant of the laser pulse and to transmit the THz wave. Three metal-mesh bandpass filters with the center frequency of $\omega=0.5$ THz were used to make the narrow-band multicycle THz pulse [S5]. We used four wire-grid polarizers (WGPs) with their angles $\theta_1$, $\theta_2$, $\theta_3$, and $\theta_4$, as described in the main text. The transmitted THz wave was recorded by the electro-optic sampling in a (100) ZnTe crystal with the balanced detection.

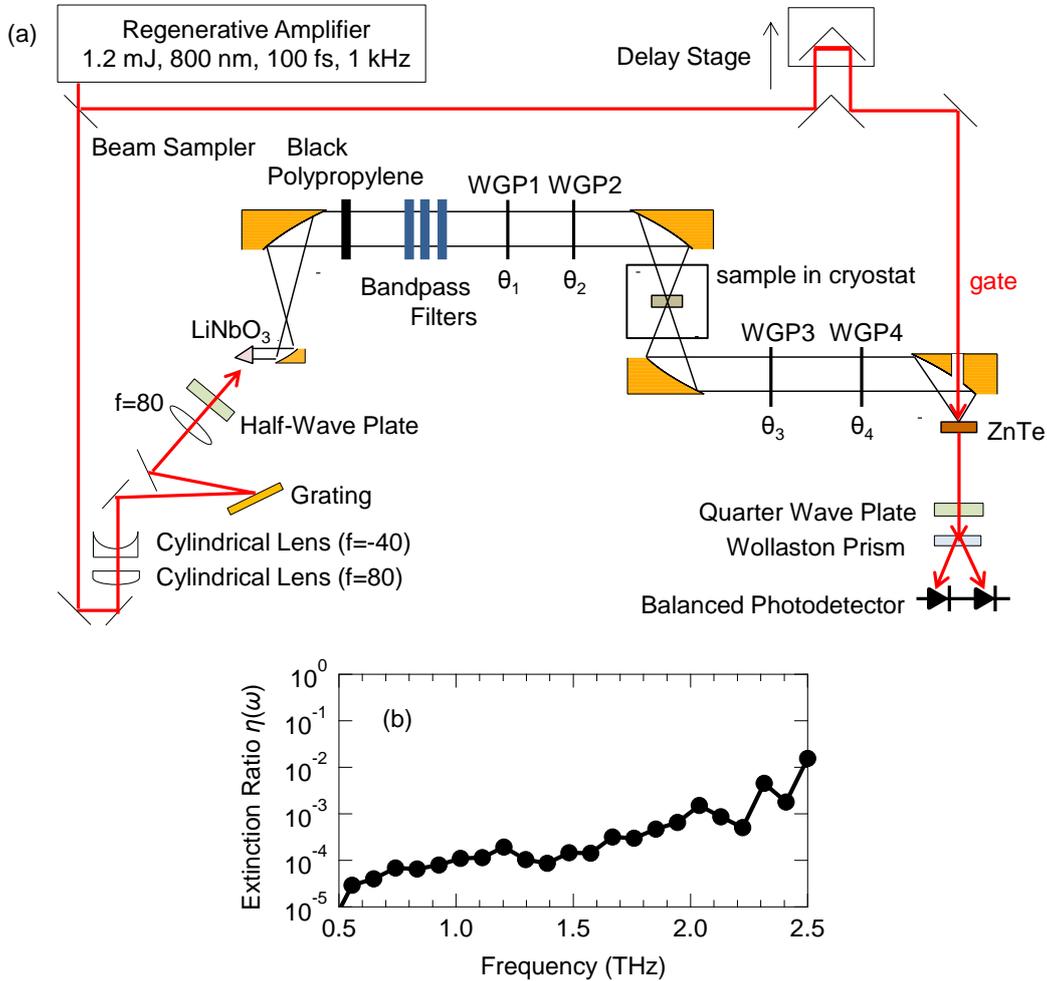

Fig. S2. (a) Schematics of the polarization-resolved THz transmission spectroscopy setup. (b) Extinction ratio $\eta(\omega)$ as a function of frequency.

The polarization resolution of the present setup was evaluated as follows. With the WGP angles of $\theta_1=\theta_2=\theta_3=0°$ and $\theta_4=45°$ (parallel configuration), we obtained the THz waveform $E_{//}(t)$ without bandpass filters. After rotating the WGP3 at $\theta_3$=90º (perpendicular configuration), we also measured the transmitted THz wave $E_\perp(t)$. Then, after performing the Fourier transformation, the extinction ratio $\eta(\omega)$ defined by

$$\eta(\omega) = \frac{|E_\perp(\omega)|^2}{|E_{//}(\omega)|^2}$$

was obtained as shown in Fig. S2(b). The extinction ratio is ~$10^{-4}$ below $3\omega$=1.5 THz, corresponding to an angle resolution of ~0.6º, which is sufficient for resolving the polarization of the fundamental and third-harmonic THz waves in our experiments. We confirmed that distortions of the polarization caused by the parabolic mirrors, the sample, and the inner and outer windows in the cryostat were negligibly small. The frequency dependence of the extinction ratio in Fig. S2(b) is attributed to the WGPs with 10-μm wire diameters and 20-μm spacings.

**Analysis and Corrections for the Third-Harmonic Generation (THG)**

To investigate the incident polarization angle ($\theta_2$) dependence of the THG intensity, we measured the transmitted intense narrow-band THz pulse with the WGP angles $\theta_3$=$\theta_2$ (parallel configuration) and $\theta_4$=45º. Power spectra of the transmitted pulses were obtained by the Fourier transform with the Blackman window function to suppress sidelobes. By integrating the power spectra from 1.3 to 1.7 THz and from 0.3 to 0.7 THz, we defined the observed THG intensity $I_{3\omega}^{obs}$ and the observed fundamental intensity $I_{\omega}^{obs}$, respectively. The observed values of $I_{3\omega}^{obs}$ and $I_{\omega}^{obs}$ are related with the generated THG intensity $I_{3\omega}(\theta_2, E_{THz})$ and the transmitted fundamental intensity $I_\omega(\theta_2, E_{THz})$, respectively, by the relations,

$$I_{3\omega}^{obs} = I_{3\omega}(\theta_2, E_{THz})\cos^2(\theta_2 - 45°),$$

$$I_\omega^{obs} = I_\omega(\theta_2, E_{THz})\cos^2(\theta_2 - 45°),$$

because of the projection on the WGP4. Then we evaluated the ratio,

$$R(\theta_2) \equiv \frac{I_{3\omega}^{obs}}{I_\omega^{obs}} = \frac{I_{3\omega}(\theta_2, E_{THz})}{I_\omega(\theta_2, E_{THz})} \propto \frac{|\chi(\theta_2)|^2 |E_{THz}|^6}{|E_{THz}|^2} = |\chi(\theta_2)|^2 |E_{THz}|^4,$$

where $\chi(\theta_2)$ is the third-order nonlinear susceptibility and $E_{THz}$ is the incident field strength. When $E_{THz}$ is kept constant during the experiment, $E_{THz}$ should be independent of $\theta_2$, so that $R(\theta_2)$ directly reflects the incident polarization angle dependence of the squared nonlinear susceptibility, $|\chi(\theta_2)|^2$. Figure S3 shows $R(\theta_2)$ against $\theta_2$, which indicates that the $\theta_2$ dependence is very small.

Note that $R(\theta_2)$ is very sensitive to the field strength $E_{THz}$. Even a slight misalignment of the WGP angles or fluctuation of the laser output during the experiment could alter $E_{THz}$, which might result in a change of $R(\theta_2)$. To check the strength of $E_{THz}$ during the polarization rotation measurements, we performed the following correction. We first assume that the $\theta_2$ dependence of the transmitted fundamental intensity $I_\omega(\theta_2, E_{THz})$ is negligibly small. This is reasonable because the material shows isotropic linear response, and the conversion efficiency to the THG is small (<$10^{-3}$) [S5]. Then the observed fundamental intensity $I_\omega^{obs}$ should follow $\cos^2(\theta_2$-45º) because of the projection on the WGP4 as long as the field strength is constant. By comparing $I_\omega^{obs}$ with $\cos^2(\theta_2$-45º), we evaluated the change of the field strength as a function of $\theta_2$ in the form

of
$$E_{\text{THz}}(\theta_2) \propto \sqrt{I_\omega^{\text{obs}}/\cos^2(\theta_2 - 45°)}.$$

With this expression we obtained the $\theta_2$ dependence of the squared nonlinear susceptibility as
$$|\chi(\theta_2)|^2 \propto R(\theta_2)/|E_{\text{THz}}(\theta_2)|^4$$

as plotted in Fig. S3 as well as Fig. 3(c) in the main text. Figure S3 shows that the effect of this correction was negligibly small and that both $R(\theta_2)$ and $|\chi(\theta_2)|^2$ are quite isotropic.

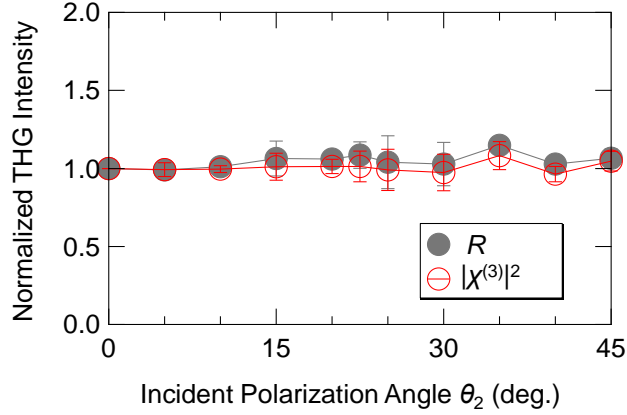

**Theoretical methods**

The THG susceptibility is evaluated within the BCS approximation and dynamical mean-field theory (DMFT). The screening effect due to the long-range Coulomb interaction is here taken into account within the random-phase approximation. We assume the approximate particle-hole symmetry which is valid in the vicinity of the Fermi energy. The general expression for the CDF contribution to the THG susceptibility is given by the multi-orbital generalization of [S6],

$$\chi_{\text{CDF}}(\omega) = \sum_{k,a}\left(\sum_{i,j}\frac{\partial^2 \varepsilon_a(k)}{\partial k_i \partial k_j}e_i^I e_j^O\right)\left(\sum_{m,n}\frac{\partial^2 \varepsilon_a(k)}{\partial k_m \partial k_n}e_m^I e_n^I\right)\chi_{33,a}(\omega,k) - \frac{\left[\sum_{k,a}\sum_{i,j}\frac{\partial^2 \varepsilon_a(k)}{\partial k_i \partial k_j}e_i^I e_j^O \chi_{33,a}(\omega,k)\right]\left[\sum_{k,a}\sum_{m,n}\frac{\partial^2 \varepsilon_a(k)}{\partial k_m \partial k_n}e_m^I e_n^I \chi_{33,a}(\omega,k)\right]}{\sum_{k,a}\chi_{33,a}(\omega,k)},$$

where $\omega$ is the frequency of the incident light, $e^I$ and $e^O$ ($|e^I|=|e^O|=1$) are the polarization vectors of the incident and transmitted light, and $\chi_{33,a}(\omega,k)$ is the bare charge-charge susceptibility for orbital $a$. Within the BCS approximation, $\chi_{33,a}(\omega,k)$ is given by

$$\chi_{33,a}(\omega,k) = \frac{\Delta^2}{2E_a(k)[\omega^2 - E_a(k)^2]}\tanh\left(\frac{E_a(k)}{2T}\right)$$

where $\Delta$ is the superconducting gap, $T$ is the temperature, and $E_a(k) = \sqrt{\varepsilon_a(k)^2 + \Delta^2}$. The divergence in $\chi_{33,a}(\omega,k)$ is regularized by replacing $\omega$ with $\omega+i\delta$ with a broadening factor $\delta=0.1$ meV. In DMFT, on the other hand, $\chi_{33,a}(\omega,k)$ is calculated via Eq. (2) in Ref. [S7], where we use $\delta=0.02$ eV as a broadening factor for the electron and phonon Green's functions. If $e^I$ and $e^O$ are parallel to the $xy$ plane, the polarization dependence is generally given by Table 1 in the main text, where $A(\omega)$ and $B(\omega)$ are given as

$$A(\omega) = \sum_{k,a}\left(\frac{\partial^2\varepsilon_a(k)}{\partial k_x^2}\right)^2 \chi_{33,a}(\omega,k) - \frac{1}{\sum_{k,a}\chi_{33,a}(\omega,k)}\left[\sum_{k,a}\frac{\partial^2\varepsilon_a(k)}{\partial k_x^2}\chi_{33,a}(\omega,k)\right]^2,$$

$$B(\omega) = \frac{1}{4}\sum_{k,a}\left[-\left(\frac{\partial^2\varepsilon_a(k)}{\partial k_x^2}\right)^2 + \frac{\partial^2\varepsilon_a(k)}{\partial k_x^2}\frac{\partial^2\varepsilon_a(k)}{\partial k_y^2} + 2\left(\frac{\partial^2\varepsilon_a(k)}{\partial k_x\partial k_y}\right)^2\right]\chi_{33,a}(\omega,k).$$

The Higgs-mode contribution for the THG susceptibility comprises three components: the non-resonant, mixed, and resonant contributions [S7]. Within the BCS approximation, the latter two identically vanish, and the remaining non-resonant contribution is given by

$$\chi_H(\omega) = \left[\sum_{k,a}\sum_{i,j}\frac{\partial^2\varepsilon_a(k)}{\partial k_i\partial k_j}e_i^I e_j^O \chi_{13,a}(\omega,k)\right]\left[\sum_{k,a}\sum_{m,n}\frac{\partial^2\varepsilon_a(k)}{\partial k_m\partial k_n}e_m^I e_n^I \chi_{13,a}(\omega,k)\right]\Gamma(\omega),$$

where $\chi_{13,a}(\omega,k)$ is the pair-charge susceptibility for orbital $a$ that depends on $k$ only through the factor $\varepsilon_a(k)$, while $\Gamma(\omega)$ is the momentum-independent vertex function for the pair amplitude [S7]. If the lattice structure has the parity symmetry ($\varepsilon_a(-k)=\varepsilon_a(k)$) as in the present case, the terms with $i \neq j$ and $m \neq n$ vanish. Since all the directions ($x$, $y$, $z$) are equivalent after the orbital and momentum summations, $\sum_{k,a}(\partial^2\varepsilon_a(k)/\partial k_i^2)\chi_{13,a}(\omega,k)$ does not depend on $i$. Hence the polarization dependence of the Higgs-mode contribution $\chi_H(\omega)$ is given as in Table 1 in the main text with

$$C(\omega) = \left[\sum_{k,a}\frac{\partial^2\varepsilon_a(k)}{\partial k_x^2}\chi_{13,a}(\omega,k)\right]^2\Gamma(\omega).$$

Namely, the polarization of the third harmonics originating from the Higgs mode is always parallel to the polarization of the incident light, and its intensity does not depend on the polarization angle of the incident light $\theta$. In DMFT, on the other hand, the mixed and resonant Higgs-mode contributions exist as well. One can show that they have the same polarization dependence as the non-resonant one from an argument similar to the above.